\documentclass[useAMS,usenatbib]{mn2e}
\usepackage{graphicx}
\usepackage{rotating}
\usepackage{aas_macros}
\usepackage{amsmath}
\usepackage{color}

\def\etal{{\it et al. }} 
\title[Globular Clusters and the Dark Matter Content of NGC~3923] 
{The Globular Cluster Kinematics and Galaxy Dark Matter Content of NGC~3923} 

\author[Norris \etal] {Mark A. Norris$^{1, 2}$\thanks{manorris@physics.unc.edu}, Karl Gebhardt$^{3}$, 
Ray M. Sharples$^{2}$, Favio Raul Faifer$^{4,5}$,\newauthor
Terry Bridges$^{6}$, Duncan A. Forbes$^{7}$, Juan C. Forte$^{5}$, Stephen E. Zepf $^{8}$,\newauthor 
Michael A. Beasley$^{9}$, David A. Hanes$^{6}$, Robert Proctor$^{10}$, Sheila J. Kannappan$^{1}$ \\
\\
  $^1$ Dept. of Physics and Astronomy UNC-Chapel Hill, CB 3255, 
  Phillips Hall, Chapel Hill, NC 27599-3255, USA \\
  $^2$ Department of Physics, University of Durham, 
  South Road, Durham, DH1 3LE, UK\\
  $^3$ Astronomy Department, University of Texas, 
  Austin, TX 78712, USA\\
  $^4$ Instituto de Astrofisica de La Plata (CCT La Plata - CONICET - UNLP) \\
  $^5$ Facultad de Cs. Astronomicas y Geofisicas, 
  UNLP, Paseo del Bosque 1900, La Plata, and CONICET, Argentina\\
  $^6$ Department of Physics, Queen's University, 
  Kingston, ON K7L 3N6, Canada\\
  $^7$ Centre for Astrophysics \& Supercomputing, Swinburne University, 
  Hawthorn, VIC 3122, Australia\\ 
   $^8$ Department of Physics and Astronomy, Michigan State University, 
  East Lansing, MI 48824, USA\\
  $^9$ Instituto de Astrofisica de Canarias, 
  La Laguna 38200, Tenerife, Spain\\ 
  $^{10}$Universidade de S\~{a}o Paulo, IAG, Rua do Mat\~{a}o, 1226, S\~{a}o Paulo, 05508-900, Brasil \\  
}

\begin{document}

\date{Accepted 2007 ***. Received 2007 ***; in original form ***}

\pagerange{\pageref{firstpage}--\pageref{lastpage}} \pubyear{2007}

\maketitle

\label{firstpage}

\begin{abstract}
This paper presents further results from our spectroscopic study of the globular cluster (GC) system of 
the group elliptical NGC~3923. From observations made with the GMOS instrument on the Gemini South telescope, an 
additional 50 GC and Ultra Compact Dwarf (UCD) candidates have been spectroscopically 
confirmed as members of the NGC~3923 system. When the recessional velocities of these GCs are 
combined with the 29 GC velocities reported previously, a total sample of 79 GC/UCD 
velocities is produced. This sample extends to over 6 arcmin ($>$6~R$_e$ $\sim$30 kpc) from the 
centre of NGC~3923, and is used to study the dynamics of the GC system and the dark matter content 
of NGC~3923. It is found that the GC system of NGC~3923 displays no appreciable rotation, and that 
the projected velocity dispersion is constant with radius within the uncertainties. The velocity dispersion 
profiles of the integrated light and GC system of NGC~3923 are indistinguishable over the region in 
which they overlap. We find some evidence that the diffuse light and GCs of NGC~3923 have radially 
biased orbits within $\sim$130$^{\prime\prime}$. The application of axisymmetric orbit-based models 
to the GC and integrated light velocity dispersion profiles demonstrates that a significant increase in 
the mass-to-light ratio (from M/L$_{\rm V}$ = 8 to 26) at large galactocentric radii is required to explain 
this observation. We therefore confirm the presence of a dark matter halo in NGC~3923. We find that 
dark matter comprises 17.5$^{+7.3}_{-4.5}$$\%$ of the mass within 1~R$_{\rm e}$, 41.2$^{+18.2}_{-10.6}$$\%$ 
within 2~R$_{\rm e}$, and 75.6$^{+15.4}_{-16.8}$$\%$ within the radius of our last kinematic tracer at 
6.9~R$_{\rm e}$. The total dynamical mass within this radius is found to be 1.5$^{+0.4}_{-0.25}$$\times$10$^{12}$~M$_\odot$.
In common with other studies of large ellipticals, we find that our derived dynamical mass profile is 
consistently higher than that derived by X-ray observations, by a factor of around 2.

\end{abstract}


\section{Introduction}
 \label{Sec:Introduction}
 
The mass distributions and dark matter (DM) content of galaxies
are of significant importance to the study of the formation and
evolution of galaxies. The current cold dark matter (CDM) paradigm 
posits that dark matter is the dominant matter component found in the 
Universe, and hence drives the formation of structure throughout
the history of the Universe. However making measurements of the actual 
distribution of DM with which to confront theory can prove to be difficult. 
In particular, measuring the DM distribution on the scales of individual galaxies can 
prove to be extremely trying for any galaxy type other than the most massive
early-type galaxies (where significant X-ray emission is present)
or later-type spirals (where gas emission can be traced to large radii to trace
flat rotation curves, e.g. \citealt{Persic96}). Intermediate to low mass 
early-type galaxies are of notable concern, as these galaxies
usually lack significant X-ray or ionised gas emission, and generally lack readily
observable tracers of the stellar kinematics at large radii (such as absorption line
kinematics), due to the inherent faintness of the stellar populations at the 
radii where DM is expected to dominate (beyond 2R$_e$).  For these galaxy 
types then, the use of dynamical probes of the galaxy potential such as GCs and 
planetary nebulae (PNe) are of particular importance.

GCs are well suited to this type of study for several reasons: (i)
their ubiquity around all relatively massive galaxies (e.g. see \cite{Peng06}
for Virgo Cluster galaxies); (ii) their extended distribution, that ensures 
significant numbers can be found beyond the radii at which the DM
component becomes significant; (iii) and the fact that they are sufficiently luminous
to be spectroscopically studied to large distances ($\sim$30Mpc). 
In addition to providing determinations of the relative amounts of dark and
baryonic matter, the study of the kinematics of large samples of GCs around 
early-type galaxies also provides other useful clues to their formation
histories. In some galaxies for example, the GC system (or a subpopulation)
is found to rotate with the galaxy (see below), perhaps indicative of a 
common formation history for the stellar component and the GC system. 

The examination of the kinematics of GC systems is therefore
an extremely promising area of research, with much to be learned
about the processes that lead to the formation of currently observed galaxies. 
In recent years, a growing number of studies have made use
of 8m class telescopes to build large samples of GC velocities around
several early-type galaxies. In general, these have been of more massive
cluster galaxies, known to harbour large GC populations. 
The project of which this study is a part intends to widen the host galaxy
mass and environment range so far studied. To date, NGC~3379
\citep{Pierce3379} and NGC~4649 \citep{Bridges06} have been studied.
Both galaxies displayed GC velocity dispersion profiles that
were constant with radius, a finding indicative of the presence
of DM. 

To these studies can be added investigations of the GC systems
of NGC~1399 \citep{Richtler08,Schuberth10}, NGC~1407 \citep{Romanowsky09}, 
NGC~4472 \citep{Sharples98,Zepf00,Cote03}, NGC~4486 \citep{Cote01,StraderM87}, 
NGC~4636 \citep{Schuberth06,Chakrabarty08,Park10}, and NGC~5128 \citep{Woodley07}
all of which have found similarly constant or slightly increasing GC velocity dispersion 
profiles, providing strong evidence for the existence of significant
amounts of DM in these galaxies. 

Examination of the GC systems of these galaxies for signatures of rotation
like that displayed by the Milky Way metal-rich GC population has 
been less conclusive. Significant rotation has been observed in some 
galaxies  (e.g. NGC~3115; \citealt{Kuntschner02, ArnoldNGC3115} and NGC~5128; \citealt{Woodley07}) 
but little or none in others (e.g. NGC~1399; \citealt{Richtler08} $\&$ \citealt{Schuberth10}, 
NGC~1407; \citealt{Romanowsky09}, NGC~3379; \citealt{Pierce3379}, 
NGC~4472; \citealt{Zepf00,Cote03}, NGC~4594; \citealt{Bridges07}, and 
NGC~4649; \citealt{Bridges06}). In the case of NGC~4486
early studies indicated significant rotation \citep{Cote01}, but more recent 
analyses \citep{StraderM87} indicate generally insignificant amounts of 
rotation. When GCs are split into blue and red subpopulations the picture 
becomes even more complicated. Generally it can be stated that the blue 
subpopulation displays a higher velocity dispersion than the red population, 
but beyond this the picture is confused with considerable variation from 
galaxy to galaxy. 

The study of the orbits of the GCs has generally been stymied by the need for
extremely large samples of GC velocities \citep[see e.g.][who estimate that
around a thousand GC velocities would be required to solve for both the 
mass distribution and orbital anisotropy simultaneously in the case of NGC~4486]{Merritt93}. 
Where fewer velocities are available, other input such as an X-ray profile can 
be used to constrain the potential, allowing the observed dispersion profile to be used to
determine the GC orbits.  In NGC~1399 and NGC~4472 the GCs display 
orbital characteristics close to isotropy, but in NGC~1407 \citep{Romanowsky09} 
and NGC~4649 \citep{Bridges06} there is some evidence of the GCs having more 
tangential orbits at certain radii.

This paper extends the study of the GC system of NGC~3923 described 
in \cite{Norris08} to include an investigation of the kinematics of the
NGC~3923 GC system. NGC~3923 is a large (R$_{\rm e}$ of $\sim$53.4 arcsec, 4.6~kpc), nearby 
(D $\sim$ 21~Mpc) elliptical galaxy \citep[$\epsilon\sim$ 0.3 - 0.4,][]{Sikkema07},
 with a prominent shell structure \citep{Malin83}, and is the brightest galaxy of an average sized group.
To date no studies have examined the GC kinematics of this
galaxy, but several have probed the kinematics of the integrated light 
of NGC~3923.
In \cite{Norris08} diffuse light kinematics extracted from the same MOS
slitlets as the GC spectra are used to confirm that the main body of 
NGC~3923 rotates along the major axis with a small amplitude of 
$\sim$30~kms$^{-1}$. This result is consistent with the work of 
\cite{Koprolin2000}, but in marginal disagreement with the work 
of \cite{Carter98} that found no rotation around the major axis but
slight rotation of amplitude $\sim$20~kms$^{-1}$ around the minor axis.
The only previous investigation to have studied the
DM content of NGC~3923 is the X-ray study of \cite{Fukazawa06},
which found that the M/L in the B-band increased from 3.5 in the inner
regions to around 15 at 18.4 kpc, a trend strongly supporting the presence
of a DM halo associated with this galaxy.



\section{Observations and Data Reduction}
In this paper we continue the study of the GC system of NGC~3923 first presented in
\cite{Norris08}. To the 29 NGC~3923 GC velocities previously determined from 
Multi-Object Spectroscopy (MOS) we add an additional 50 GC velocities 
measured using a combination of MOS and Nod-and-Shuffle (N$\&$S) observations.

As described in \cite{Norris08}, pre-imaging for GC candidate selection was undertaken 
for 3 fields (central, SW and NE) on 2004 January 19, comprising 4$\times$200 
seconds in Sloan g, and 4$\times$100 seconds in r and i. Reduction of the
imaging data involved bias subtraction, flat fielding, fringe removal (for i images),
combining dithered sub-exposures to remove chip gaps, and finally removal
of a median filter image (to remove the galaxy diffuse light). More details of
the imaging reduction and object selection can be found in \cite{Forbes04} 
and \cite{Bridges06}, with a full analysis of the imaging data from 
all galaxies studied in this project presented in \cite{Faifer11}.

A full description of the reduction and analysis of the deep MOS
observations of the central pointing is provided in \cite{Norris08}. We present here 
data from an additional shallower ($\sim$3hrs vs. 8hrs) MOS observation centred on
NGC~3923 and three further N$\&$S observations. The shallower MOS
observation was primarily focussed on studying the internal kinematics and 
stellar populations of a sample of UCDs discovered by \citet{Norris&Kannappan11}.
This MOS mask contained 33 slitlets, with those on UCD candidates being
0.5" wide and those on the fainter GC candidates 1.5" wide. The B1200 grating
was employed providing resolution of $\sim$ 1.2 and 3.6 \AA\, and coverage
of 1490 \AA\, generally centred around 5000\AA\, (the coverage varies slightly based
on the position of the slit on the mask). These newer MOS observations were 
reduced using the same tasks and procedures as outlined in \cite{Norris08}, and 
interested readers are referred there for further details. Hereafter we shall focus 
on describing the reduction of the N$\&$S spectroscopy.

GMOS N$\&$S masks were produced for each of the 3 fields of pre-imaging. 
The central field containing the largest number of targets was therefore studied 
three times, twice with MOS spectroscopy and once with N$\&$S. The N$\&$S masks contained 
a total of 44 (central field), 46 (SW field) and 48 (NE field) slitlets, each 
1 arcsecond wide by at least 2.25 arcseconds long. The majority of the 
slitlets per mask (30-35) were placed on GC candidates, with the remaining slitlets 
placed on other objects of interest to fill up the mask. Each N$\&$S mask was
exposed using the B600$\_$G5303 grism for 4$\times$1800 seconds at a central
wavelength of 505 nm and 4$\times$1800 seconds at a central wavelength of 510 nm, 
yielding 4 hours of on-source integration per mask. The spectra were
exposed through the g$\_$G0325 and GG455$\_$G0329 filters to restrict the
wavelength coverage to a well-defined bandpass of $\sim$4600-5550\AA.
This restriction in wavelength coverage allowed more freedom in the positioning
of slitlets and hence in choice of targets, as spectra could be aligned in the 
spatial dimension without overlapping. The wavelength coverage of each spectrum 
was still sufficient to cover most of the strong absorption lines found in optical spectra 
and includes the most important lines for such studies, H$\beta$, Mg$b$, Fe5270 
and Fe5335. Bias frames, flat fields and copper-argon (CuAr) arc spectra were
observed throughout the observations as part of the standard Gemini baseline 
calibrations.

The reduction of the N$\&$S spectra was accomplished using the standard
Gemini/GMOS packages in IRAF, a brief description of which shall be given
here to help illustrate the benefits and problems associated with using this
technique for GC studies. In N$\&$S spectroscopy the target is placed
at one end of the slit, it is exposed as normal for a period of time, and
the telescope is then nodded to a sky position whilst shuffling the charge
on the CCD between science and un-illuminated storage regions. The image that
results contains two spectra - one of the object and one of the sky, both of
which importantly were obtained on exactly the same pixels after traversing
identical optical paths. Therefore, when subtracting the sky spectrum from the
object spectrum, effects such as flat-fielding errors, fringing, and variation in the sky
cancel out more accurately than in standard sky subtraction procedures. In the 
case of the data here, a small offset was applied in the nodding procedure 
so that the object appears in both the science and the sky spectra, only at 
different ends of the slit. To produce these spectra, the standard reductions for 
Gemini/GMOS IRAF N$\&$S tasks were used to provide bias subtraction, flat 
fielding, extraction of each spectra pair into separate extensions, and 
wavelength calibration. At this point it is possible to shift one spectrum up to 
overlap with the other, so that the sky region of one spectrum is over the 
target spectrum of the other. Subtracting one spectrum from the other then 
produces a sky subtracted target spectrum pair, where one spectrum is negative 
and the other positive. It is then a simple matter to trace and extract each spectrum 
as normal and combine each exposure together (remembering to flip the negative 
spectra to positive ones) to produce the final 1D spectrum.
  
In theory, the use of the N$\&$S method should have one major advantage over
standard MOS observations: the ability to pack more objects onto each mask. 
This is because the improved sky-subtraction accuracy relative to MOS observations 
means that shorter slitlets are required to determine equivalently accurate sky spectra. 
In the specific case of the data presented here, this increase in efficiency was not
achieved due to significant issues with the GMOS instrument in N$\&$S mode.
In particular the GMOS CCDs exhibit small regions that do not allow the charge
to move freely through them during the ``shuffling" part of the N$\&$S procedure.
These ``charge traps" cause trails of charge deficit. The net effect of these charge 
traps is that they lead to artifacts in the final extracted 1D spectra that are 
indistinguishable from absorption features. Several methods have been developed 
to deal with these traps (see for example \citealt{Abraham04}). However, no method is 
straightforwardly applicable to the data available here, as they generally require 
additional steps during the observations. Hence, these charge traps make 
some regions of the GMOS CCD unusable in the present case, leading to the 
unfortunate loss of up to 25$\%$ ($\sim$10 objects) of the spectra per mask.


\subsection{GC Velocity Determination}
As in \cite{Norris08}, the recessional velocities of the GC candidates
were measured using the FXCOR task in the RV package of
IRAF. This task carries out a fourier cross correlation between the
object spectra and a spectral library. In this case no radial velocity 
standard stars were observed as part of this project, 
therefore theoretical template spectra from the simple stellar population library of 
\cite{Vazdekis99} were used as templates. As described in 
\cite{Norris08} a further 6 stellar spectra from the 
\cite{Jones97} library were added to ensure adequate 
coverage of the low-metallicity region. Hence the template spectra spanned
a range in metallicity from [Fe/H] $<$ -1.5 to +0.2 and in age from 1-18 Gyr.
The quoted velocity for each object is the 3-$\sigma$ clipped mean of the 
velocities derived from the FXCOR fits to each template; the errors are estimated
from the mean of the errors measured by FXCOR for those velocities not
clipped. Only GC candidates where at least half of the templates returned
Tonry and Davis R value greater than 5 and velocities within 3-$\sigma$ of the mean 
are considered further. This ensures that only reliable
velocities are included in our analysis, and is a necessary step since only a few 
erroneous velocities can lead to significant changes in the inferred properties
such as velocity dispersion \cite[see e.g.][]{StraderM87}.


\begin{figure} 
   \centering
   \begin{turn}{0}
   \includegraphics[scale=0.7]{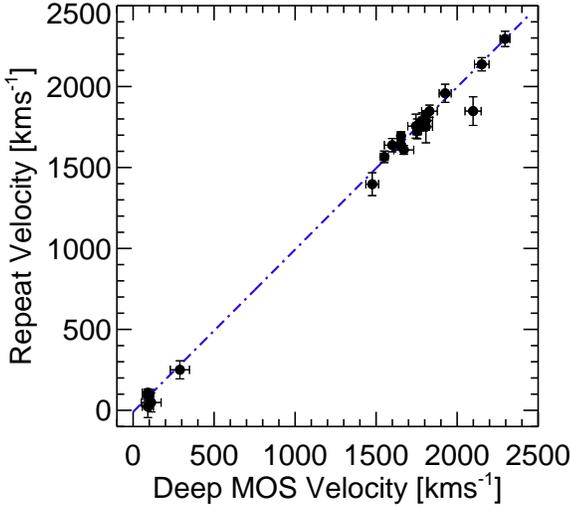}
   \end{turn}
    \caption[Comparison of Deep MOS and other Velocities]
    {Comparison of the velocities of 24 objects observed in the
    deep MOS mask and then re-observed on at least one of the shallow MOS 
    or Nod-and-Shuffle fields. 
    The dashed blue line is the best fit least squares linear relation which has the form
    Y = X$\times$1.003 ($\pm$ 0.017) - 9.79 ($\pm$26.62). The RMS scatter
    about the one-to-one relation is 62.3 kms$^{-1}$.}
   \label{fig:ngc3923_mos_ns}
\end{figure}


Objects with velocities in the range 1100-2500kms$^{-1}$ 
($\sim$V$_{\rm{gal}}$ $\pm$ 700 kms$^{-1}$, or  $\sim$3-$\sigma$ of the
GC velocity dispersion) are assumed to be associated with NGC~3923. 
This is marginally different to that quoted in \cite{Norris08}, where 1200-2400 kms$^{-1}$ 
was used. In practice, however, this makes no difference to the selection of
objects classified as GCs in \cite{Norris08}, as no additional objects
from the deep MOS observations would fall into this wider velocity range. The wider
velocity range does however allow one additional N$\&$S object to 
be included in the present sample. 


A total of 24 objects were examined with both the deep MOS and shallow 
MOS or N$\&$S 
allowing an examination of the robustness of the measured velocities and their errors. 
Figure \ref{fig:ngc3923_mos_ns}
displays the velocity measured using the shallow MOS or N$\&$S observations, versus
the velocity measured using the deep MOS observations for 24 objects that were observed 
twice. As can be seen, the agreement between the two measurements is exceptionally good, 
with the best fit linear relation between the two sets of observations being consistent with no 
offset between the two methods. This good agreement allows the simple weighted averaging 
of velocities measured using both techniques in the case of overlaps. 

\begin{figure*} 
   \centering
   \begin{turn}{0}
   \includegraphics[scale=1.0]{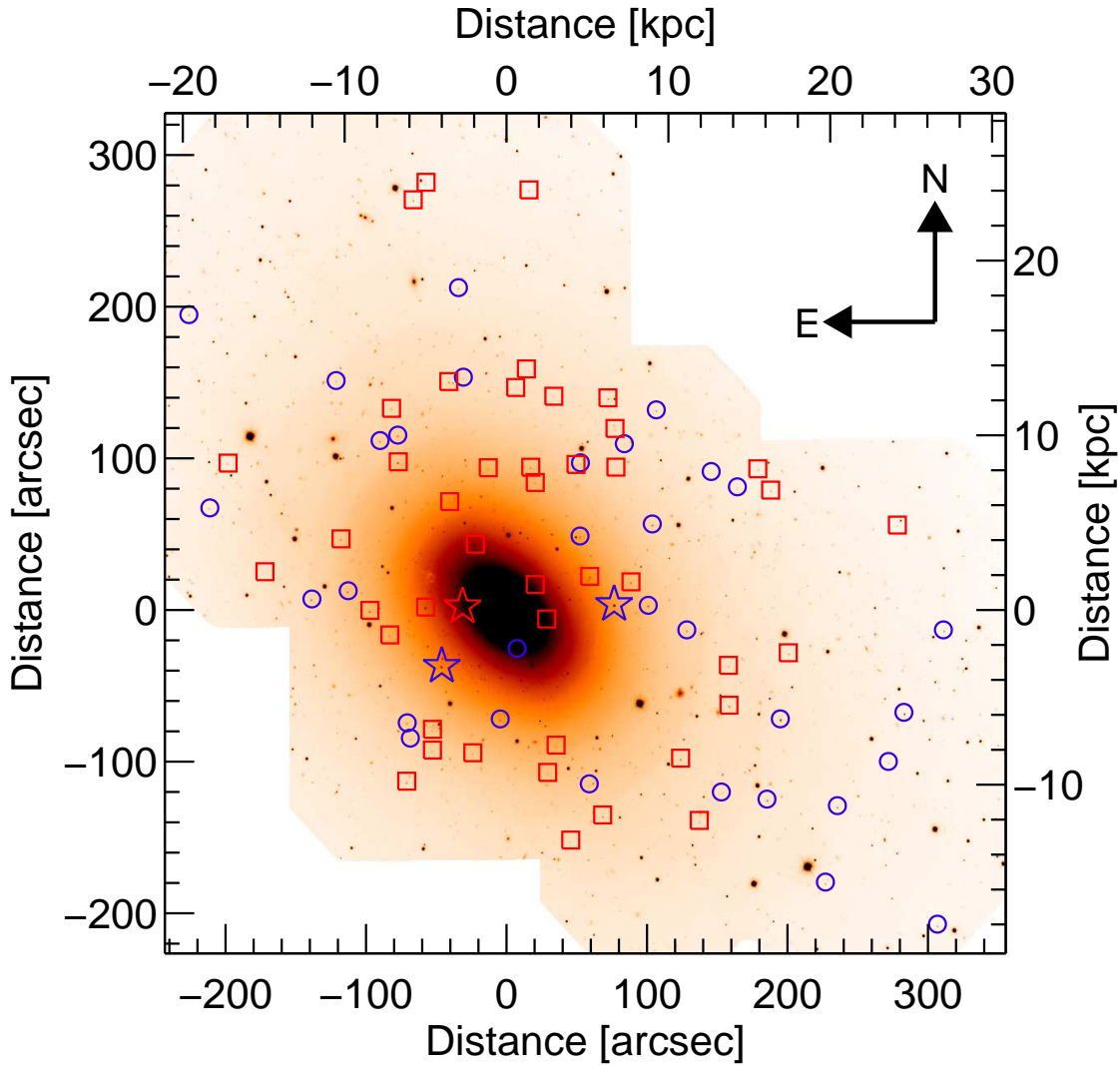}
   \end{turn}
    \caption[Confirmed NGC~3923 GCs]
    {The positions of kinematically confirmed GCs/UCDs associated
    with NGC~3923. Blue circles and red squares indicate blue and red
    GC subpopulation members, respectively. The three stars indicate the
    positions of the three confirmed UCDs discovered by \cite{Norris&Kannappan11}.
     The subpopulations were
    split at g$^\prime_0$-i$^\prime_0$ = 0.89. The physical scale is indicated based on our adopted 
    distance for NGC~3923 of 17.9 Mpc (chosen to match that used in 
    \citealt{Fukazawa06}).}
   \label{fig:ngc3923_fov}
\end{figure*}

In the few cases where N$\&$S targets were observed on two
separate N$\&$S masks but no velocity was recovered, the spectra were 
coadded and then velocity measurement reattempted. However, in no case were
any additional velocities recovered. In total, after combination of repeat observations,
79 secure velocities were determined. The spatial distribution of 
these 79 GCs and UCDs, separated into blue and red subpopulations as described in 
Section \ref{Sec:photometry}, can be seen in Figure \ref{fig:ngc3923_fov}. 
Table \ref{tab:ngc3923gckin} provides the positions, photometry \citep[from][]{Faifer11}
and velocities for the 79 objects confirmed as NGC~3923 GCs.


\section{Results}

\subsection{GC Colours}
\label{Sec:photometry}


\begin{figure} 
   \centering
   \begin{turn}{0}
   \includegraphics[scale=1.0]{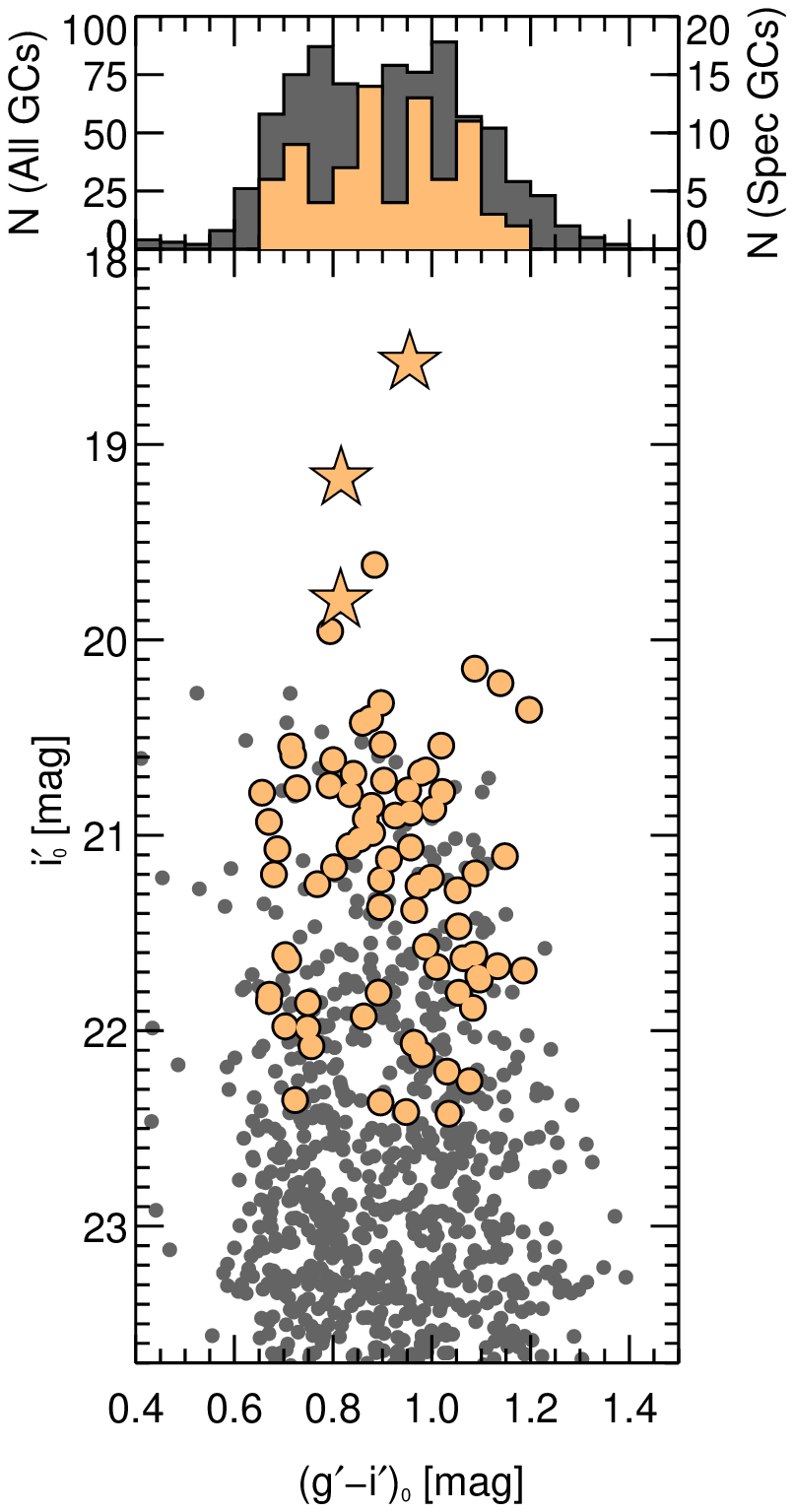}
   \end{turn}
    \caption[NGC~3923 GC Colour-Magnitude Diagram]
    {\textbf{Lower~Panel:} Colour-magnitude diagram for photometrically selected 
    GC/UCD candidates \citep[grey dots, from][]{Faifer11}, and spectroscopically confirmed
    GCs/UCDs (orange circles, this work). The three UCDs discussed in \cite{Norris&Kannappan11}
    are indicated as the orange stars. Several other objects are likely UCDs, but lie outside
    the region covered by HST imaging. This makes it impossible to definitively prove that their 
    half light radii meets the necessary limit to be classified as UCDs (R$_{\rm e}$ $>$ 6-8pc).
     \textbf{Upper~Panel:} Histogram of GC/UCD 
    g$^\prime_0$-i$^\prime_0$ colours for all GC/UCD candidates (grey histogram, left axis), 
    and spectroscopically confirmed GCs/UCDs (orange histogram, right axis).}
   \label{fig:ngc3923_cm}
\end{figure}


In this subsection we examine the photometry of the NGC~3923 GCs in order to
be able to robustly separate the GCs into blue and red subpopulations. 
Figure \ref{fig:ngc3923_cm} displays the colour-magnitude diagram for all candidate
GCs/UCDs detected in the GMOS preimaging (grey dots), as found by \citet{Faifer11}.
Additionally, the 79 spectroscopically confirmed GCs/UCDs are overplotted as large
filled orange circles/stars. In the upper panel of Figure \ref{fig:ngc3923_cm} the respective
histograms of g-i colour for each of the two samples (all potential GCs/UCDs and spectroscopically
confirmed GCs/UCDs) are also displayed. 

It can be seen from Figure \ref{fig:ngc3923_cm}
that a significant fraction of the brightest GC candidates were not spectroscopically
observed. This is largely due to problems placing pairs of slitlets because of potential 
spectral overlaps. Other potential GC candidates could not be observed or were
subsequently lost because the GCs were located too close to the galaxy 
centre, or because the spectra were unusable due to the previously described
problems with the N$\&$S observations.

The histogram of the full GC candidate sample displays a dip at around g$^\prime_0$-i$^\prime_0$ $\sim$ 0.87 
characteristic of the existence of a bimodal colour distribution. In their comprehensive 
analysis of the GC photometry \citet{Faifer11} find unambiguous bimodality
in the colours of the GCs of NGC~3923 (with a dividing colour of g$^\prime_0$-i$^\prime_0$ = 0.89), in line with 
other studies of NGC~3923 using both ground based \citep{Zepf95}, and \textit{HST} 
photometry \citep{Sikkema06,ChoThesis,Norris&Kannappan11}. Importantly however, 
\citet{Faifer11} find that bimodality is strongest for the fainter GCs and disappears 
for the brightest GCs. This result can be interpreted as evidence for a blue tilt 
\citep[e.g.][]{Strader06,Mieske06,Harris06,Peng09,Forbes10b} in the colours of the 
blue GC population of NGC~3923, where the disappearing bimodality is simply due to 
the merging of the blue GCs into the red locus at high luminosities. 


The presence or absence of a blue tilt in the NGC~3923 GC system is important.
As it is only possible to measure velocities for the most luminous GCs  (see 
Figure~\ref{fig:ngc3923_cm}), significant numbers of ``red" GCs may actually 
have formed as part of the blue subpopulation. Therefore, studies of the kinematics of 
GC subpopulations may suffer significant bias in their derived properties, due to 
contamination of the red population by members of the blue population. Unfortunately,
with the relatively limited dataset on hand, it is not possible to carry out a more 
sophisticated separation of blue and red population GCs (perhaps by allowing a
sloping separation in colour-magnitude at higher luminosity). Consequently, in all
analyses to follow, where necessary the GCs/UCDs are separated using g-i = 0.89. This
separation divides our total sample of 79 objects into subsamples of 34 blue and 
45 red objects.

\subsection{GC Spatial Distributions}
\label{Sec:spatial_dist}


\begin{figure} 
   \centering
   \begin{turn}{0}
   \includegraphics[scale=0.9]{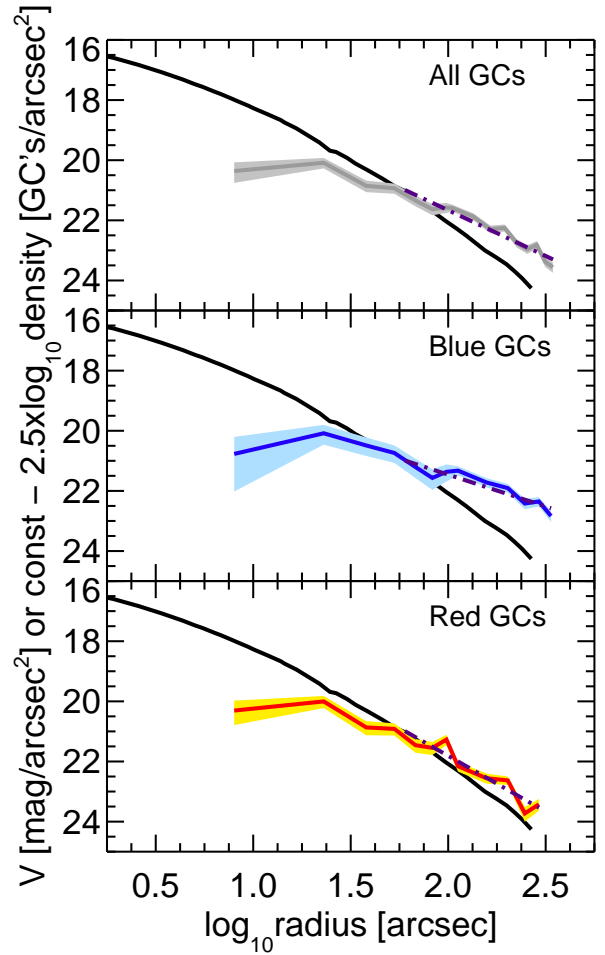}
   \end{turn}
    \caption[GC surface density profiles]
    {GC surface density profiles. Solid dark lines in each panel are the galaxy
    V band surface brightness profile. Solid coloured lines in each panel (grey, blue and red) show 
    the GC surface density profiles for the full GC sample, the blue subpopulation
    and the red subpopulation respectively. The profiles have been normalised to match
    the galaxy starlight profile at 60 arcseconds. The number counts of GCs within this
    region are expected to suffer significant incompleteness due to the high background
    \citep[see][]{Faifer11}. Shaded regions show the 1-$\sigma$ 
    poisson error regions. The purple dot-dashed line in each plot shows the best
    fit to the GC surface density profiles for GCs with galactocentric radii $>$ 60 arcseconds.
    }
   \label{fig:surface_brightness}
\end{figure}


The spatial distributions of GC subpopulations provide useful clues to
their formation, as well as being essential inputs into the dynamical models
to be described in Section \ref{Sec:dynamical_models}.

Figure \ref{fig:surface_brightness} displays the GC surface density profiles
of the total, blue, and red samples for the full photometrically selected sample
from \citet{Faifer11}. The profiles were constructed by normalising GC number
counts from \textit{HST} ACS imaging of the inner regions (because it had higher
completeness than the GMOS imaging in the inner region) to the wider field 
GMOS imaging in an overlap region of radius 60 to 90 arcsecs. The profiles
were determined in circular annuli, as no significant evidence of ellipticity 
in the GC system was found. They were also corrected for areal incompleteness,
to take into account the arrangement of the GMOS imaging pointings. In each
panel of Figure  \ref{fig:surface_brightness} the profiles have been normalised 
to match the NGC~3923 V-band surface brightness profile at 60 arcsec, allowing 
direct comparison between the GC and NGC~3923 diffuse light distributions.

It can be seen that the red GCs are more centrally concentrated and have a steeper 
number density profile than that of the blue GCs, a result common to many studies of 
the GC systems of early-type galaxies \citep[see e.g.][]{Lee1998,Cote01,Forbes04,Tamura06,Harris09b,Forte11}.
The red GC distribution is also similar to that of the diffuse light of NGC~3923,
again a common finding, at least for the innermost regions of early-types
\citep{Forbes04,Tamura06,Lee08,Bassino08}. The slight remaining offset in slope 
between the red GCs and the diffuse light of NGC~3923 can most likely be explained 
as being due to interloping blue GCs that act to reduce the red GC slope. 

The effect of contamination of one GC subpopulation by the other has not been discussed 
in great detail in the literature. However, \cite{Forte07} do discuss the possibility of 
contamination of the \textit{blue} population between the blue peak and colour valley, 
by red GCs, in the galaxies NGC~1399 and NGC~4486.

The mixing of blue GCs into the red sample due to the blue tilt is also
likely at work in the observation by \cite{Sikkema06} that the colour profile
of the red GCs is similar to that of the diffuse light of NGC~3923, but is offset 
toward the blue by around 0.05 mag. This interpretation is supported by
the results of \citet{Faifer11}, where the use of Gemini/GMOS
g and i photometry allows a better separation of the two populations, resulting
in a red GC colour profile that exactly matches that of the NGC~3923 galaxy 
halo light. Such a close correspondence between the colours
of the red GCs and the underlying integrated light is not unexpected. In
\cite{Norris08} the stellar population parameters of the diffuse light of NGC~3923 (at large radii) 
are found to be indistinguishable from those of the red GCs. This result
is consistent with the case of the edge-on S0 NGC~3115, where the stellar populations of the
bulge component at $\sim$2R$_{\rm e}$ (measured along the minor axis),
are likewise indistinguishable from those of the red GCs \citep{NSK06}.

\subsection{Kinematic Analysis}
\begin{figure} 
   \centering
   \begin{turn}{0}
   \includegraphics[scale=0.85]{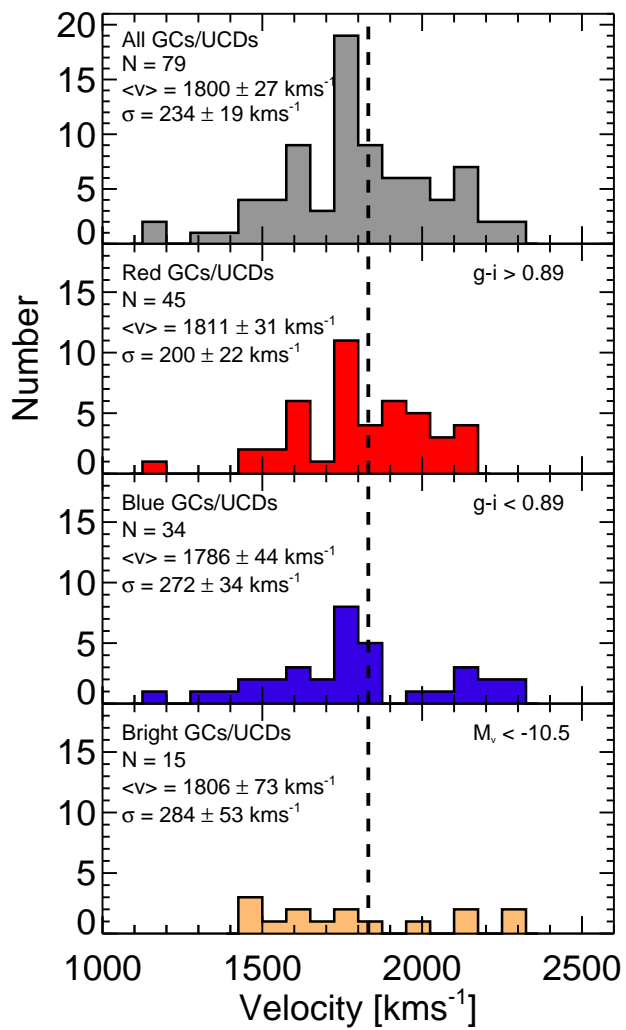}
   \end{turn}
    \caption[Histogram of GC Velocities]
    {Histogram of GC velocities, upper panel shows all 79 GCs/UCDs studied here, next panel 
    shows the 45 velocities of the red objects, the second from bottom panel is for the 35 blue objects. 
    The lowest panel shows the velocities of all objects which have M$_{\rm V}$ $<$ -10.5, roughly
    the magnitude limit of the onset of UCD-like behaviour \citep[see e.g. ][]{Norris&Kannappan11}.
    Of these 15 objects, 3 are confirmed UCDs, 2 are high probability UCD candidates (being partially
    resolved in our GMOS imaging) and several others are likely candidates requiring HST size
    determinations for confirmation.
    The vertical dashed line in all panels denotes the systemic velocity (1832 kms$^{-1}$ \citealt{Carter98}) 
    of NGC~3923.}
   \label{fig:ngc3923_hist}
\end{figure}

Figure \ref{fig:ngc3923_hist} shows a histogram of the velocities of all 79 confirmed GCs/UCDs, 
and for the blue (34), red (45), and brightest (15 objects with M$_{\rm V}$ $<$ -10.5) subpopulations separately, 
divided as described in Section~\ref{Sec:photometry}. As can be seen from Figure \ref{fig:ngc3923_hist}, 
unlike in our studies of NGC~4649 \citep{Bridges06} or NGC~3379 \citep{Pierce3379}, the 
recessional velocity of NGC~3923 ($\sim$1800 kms$^{-1}$) is sufficiently high that
contamination from Milky Way stars can be ignored. An examination of the environment 
around NGC~3923 also demonstrates that contamination of the sample by GCs associated 
with nearby neighbour galaxies is likely to be negligible. Therefore it is assumed that all 
79 objects with velocities between 1100 and 2500 kms$^{-1}$ are GCs/UCDs associated with 
NGC~3923.

We choose to include in our sample the three UCDs discovered around NGC~3923 by 
\cite{Norris&Kannappan11} (including NGC3923-UCD3 which we spectroscopically confirm here). We do this
because the preponderance of the evidence is that these UCDs formed alongside the GCs of 
NGC~3923, and are merely the most massive members of the NGC~3923 GC system.
The evidence for this is compelling; the UCDs luminosities and frequencies are consistent with being  
members of the NGC~3923 GCLF, the multiplicity of UCDs (at least 3, with 2 further high probability
candidates just within the coverage of our GMOS imaging) makes
a stripping origin unlikely, and most importantly their stellar populations are indistinguishable from
those of the GCs of NGC~3923 (\citealt{Norris&Kannappan11}, Norris et al. in prep).


Overplotted on Figure \ref{fig:ngc3923_hist} are the mean velocity
and velocity dispersion of the various GC samples. The mean velocities
of the blue and red subsamples differ by 25 kms$^{-1}$, however,
this difference is not statistically significant given the 54 kms$^{-1}$
uncertainty, and likely reflects small number statistics. Likewise,
while the blue GCs appear to have a higher dispersion than the reds (the difference is
72 $\pm$ 40 kms$^{-1}$), this result is also not significant given the present sample size.
The observation that the two GC subpopulations have similar, or only marginally different
velocity dispersion, is common 
to many studies of early-type GC kinematics \citep[see e.g.][]{Beasley04,Richtler04,Woodley10}.
Nevertheless, if increased statistics were to show that the blue GCs do
indeed display a higher velocity dispersion this would not be unexpected,
NGC~1399 \citep{Schuberth10}, NGC~4472 \citep{Zepf00,Cote03} and 
NGC~4636 \citep{Park10} all have blue GC subpopulations which are
kinematically hotter than the red subpopulations.
A higher velocity dispersion for the blue clusters could be explained naturally if some of 
the blue population are actually intra-cluster/intra-group
GCs, or were accreted from dwarf galaxies stripped by the current host.

Most interesting is the behaviour of the most luminous NGC~3923
GCs and UCDs. The bottom panel of Figure \ref{fig:ngc3923_hist} demonstrates
that unlike the case for the full sample, the most luminous objects show no
preference for velocities close to the systemic velocity \citep[similar to the behaviour
seen in the M87 GC/UCD system by][]{StraderM87}. This observation can
have two possible explanations: 1) That the luminous GC population is
significantly contaminated by objects which are not NGC~3923 GCs, such as
UCDs formed by the stripping of dwarf galaxy nuclei. 2) That the most luminous
GCs have orbits which differ from less luminous GCs, perhaps due to the preferential
destruction through dynamical friction of massive (and more extended) GCs on orbits which bring 
them close to the centre of NGC~3923. Given the small numbers currently on
hand it is hard to definitively prove which explanation is the correct one.
However, we believe that 1) is unlikely, given the fact that it would
likely require a significant fraction of the luminous NGC~3923 GCs to be
interlopers.

In the following analysis, NGC~3923 is assumed to have
an effective radius (R$_{\rm e}$) of 53.4 arcsec, as reported by \cite{Fukazawa06} 
and originally determined by \cite{Faber89} in the B band. The use of this value, which 
differs from the value of 43.8 arcseconds (as measured in the J-band) quoted in \cite{Norris08},
allows consistency between the present study and the X-ray one of \cite{Fukazawa06}.
Therefore, using this value of R$_{\rm e}$ our last measured tracer is located at $\sim$6.9 R$_{\rm e}$.
The measurement of GC kinematics at such large radii is important,
because at these radii the mass of the stellar component should be
negligible. This makes the detection of a DM component much simpler
than in the inner regions of galaxies, where the baryonic component
dominates.

\begin{figure} 
   \centering
   \begin{turn}{0}
   \includegraphics[scale=0.825]{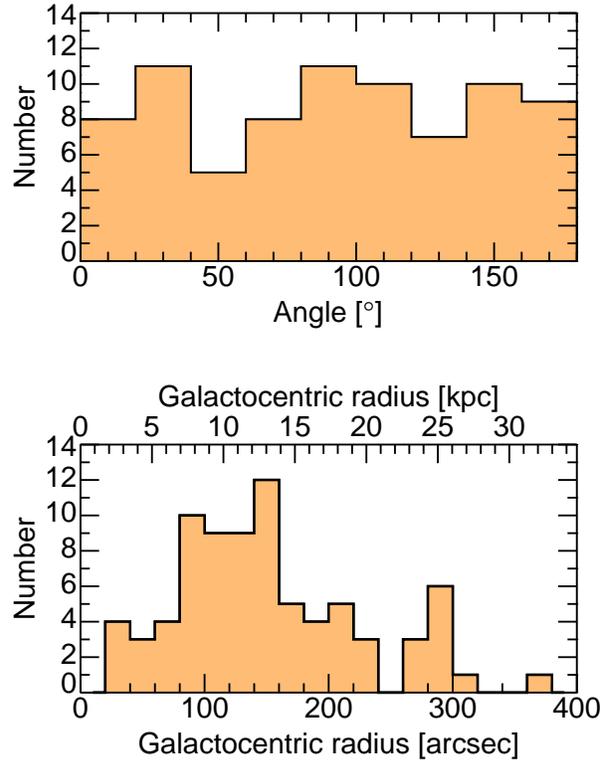}
   \end{turn}
    \caption[Histograms of Azimuthal Angle and Radius]
    {$\bf{Upper~Panel:}$ Histogram of azimuthal angle (measured North-through-East
    and folded about 180$^\circ$). $\bf{Lower~Panel:}$
    Histogram of galactocentric radius for the 79 NGC~3923 GCs. The major
    axis position angle of NGC~3923 is 48$^\circ$.}
   \label{fig:ngc3923_gc_hist}
\end{figure}

Figure \ref{fig:ngc3923_gc_hist} displays histograms that highlight
the distributions of azimuthal angle and galactocentric radius of
the 79 GCs. As expected from examination of Figure \ref{fig:ngc3923_fov}, 
there is a slight dearth of kinematically confirmed GCs located directly 
along the major axis of NGC~3923 (located at PA = 48$^\circ$).
It appears that this deficit is due to the combination of a slight
underdensity of candidate GCs in this region \citep[as seen in Fig. 1 of][]{Faifer11}
combined with the previously discussed problems of placing slitlets
due to overlaps. The group of objects located at around 0$^\circ$ would
produce spectra which overlapped with any placed at around 50$^\circ$. 
The histogram of galactocentric distance demonstrates that velocities have 
been measured for GCs located at large galactocentric radii, up to 
370 arcsec.


\subsection{Rotation of the GC system of NGC~3923}

\begin{figure} 
   \centering
   \begin{turn}{0}
   \includegraphics[scale=0.825]{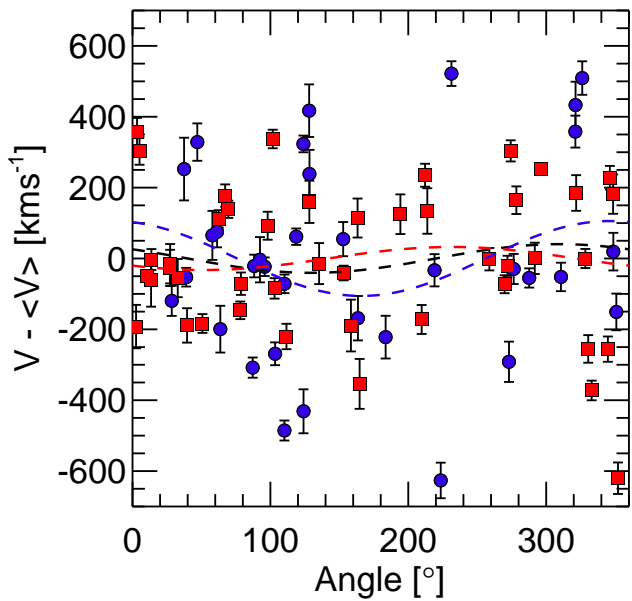}
   \end{turn}
    \caption[Velocity vs Azimuthal Angle]
    {Velocity versus azimuthal angle for all 79 NGC~3923 GCs. Blue 
    circles are blue GCs, red squares are red GCs. The black dashed line
    represents the best-fitting rotation curve for the full sample, the 
    blue and red dashed lines are the best-fitting rotation curves for the
    red and blue subpopulations. The implied rotation amplitudes are:
    32.6$\pm$44.0 kms$^{-1}$ (red), 105.7$\pm$73.3 kms$^{-1}$ 
    (blue) and 40.4$\pm$37.9 kms$^{-1}$(total), none are statistically significant.
    The major axis position angle of NGC~3923 is 48$^\circ$.}
   \label{fig:ngc3923_gc_rot}
\end{figure}

The rotation of the GC system of NGC~3923 has been examined using the procedure 
outlined in \cite{Norris08}. A non-linear least squares fit was carried out to the equation:
\\
\begin{eqnarray}
V(\theta) = V_{\rm{rot}}\rm sin(\theta - \theta_{0}) + V_{0}
\end{eqnarray}
\\
where V$_{\rm{rot}}$ is the amplitude of the GC rotation, $\theta$ is the azimuthal angle, 
$\theta_{0}$ is the position angle of the line of nodes, and V$_0$  is the recessional velocity 
of NGC~3923. This procedure was carried out for each of the three GC samples; total, red 
subpopulation, and blue subpopulation. In practice the value of V$_{0}$ was allowed to vary 
between samples. However the overall effect of this was small, as the mean velocities of the 
three groups varied by less than 10 kms$^{-1}$. The GC mean velocity (1802 $\pm$ 17 kms$^{-1}$) and
systemic velocity of NGC~3923 as determined by \cite{Carter98} (1832 $\pm$ 3 kms$^{-1}$) 
differ by 30 kms$^{-1}$. This means that using the GC mean velocity instead of the systemic
velocity could be masking some rotation. To assess whether or not this effect is significant, 
the rotation of the total sample was re-estimated using the mean GC velocity $\pm$30 kms$^{-1}$, 
in neither case was the result significantly changed. Figure \ref{fig:ngc3923_gc_rot} shows 
the best fit rotation curves derived using this procedure, in all three cases no significant 
detection of rotation is made.

The lack of rotation observed in the GC system is consistent with the
similarly small amount of rotation found for the integrated light of NGC~3923, either
found using velocities determined from the deep MOS observations presented
in \cite{Norris08}, or in smaller radii longslit studies by \cite{Carter98} (which extended
to 40 arcsec) and \cite{Koprolin2000} (which extended to 25 arcsec).
A lack of significant rotation is also a commonly observed feature of the GC systems 
of large early-type galaxies, as discussed in Section \ref{Sec:Introduction}.

The lack of any significant differences in the observed kinematics
of the GC subpopulations, and the high probability of contamination
of the ``red" GCs by blue subpopulations members lead us to focus on
the total GC population in all that follows.


\subsection{Velocity Dispersion}
\label{Sec:GCsigma}

\begin{figure*} 
   \centering
   \begin{turn}{0}
   \includegraphics[scale=0.8]{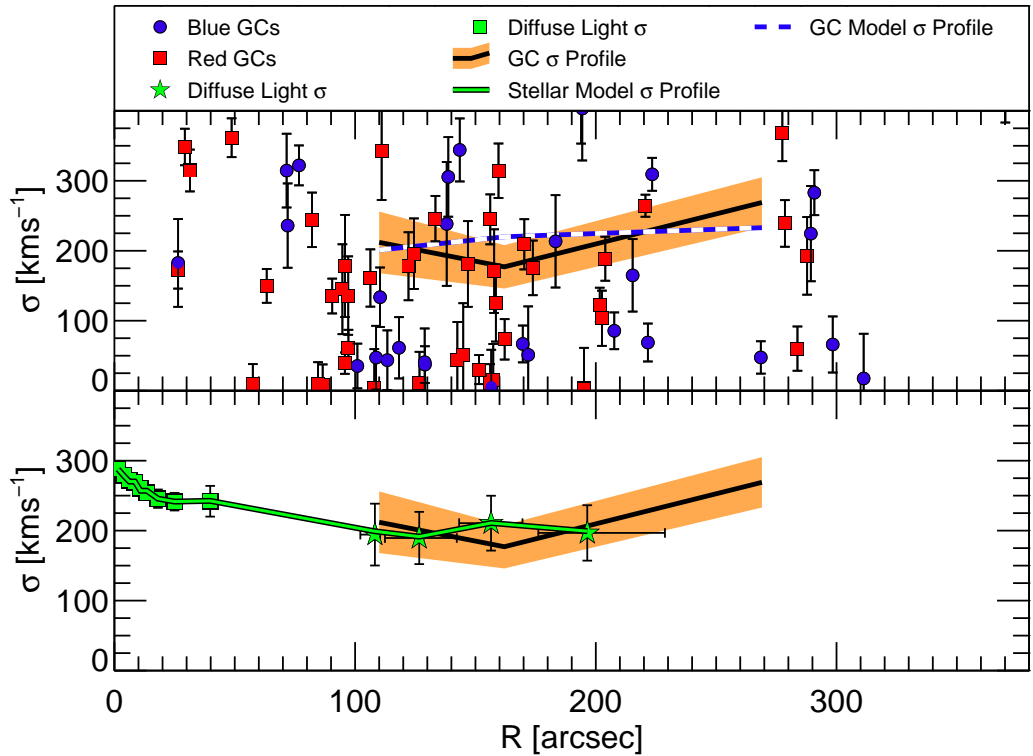}
   \end{turn}
    \caption[GC and Stellar Velocity Dispersions]
    {$\textbf{Upper~Panel:}$ Velocity dispersion versus radius for NGC~3923 GCs. Blue circles and red 
    squares are the individual blue and red GC velocities, the black solid line shows the GC dispersion
    profile estimated using the lowess method described in Section \ref{Sec:GCsigma}, the orange shaded 
    region shows the 1-$\sigma$ uncertainties on the GC velocity dispersion profile. The blue and white 
    line shows the dynamical model fit to the GC velocity dispersion profile, as described in Section 
    \ref{Sec:dynamical_models}.
    $\textbf{Lower~Panel:}$ This panel provides the comparable diffuse light measurements to those 
    provided in the upper panel. The green squares are the velocity dispersions measured for the inner 
    regions of NGC~3923 by \cite{Carter98}, the green stars represents the velocity dispersion measurements 
    of the spheroid of NGC~3923 made using the method presented in \cite{Norris08}. The solid green line 
    shows the result of fitting a dynamical model to the NGC~3923 diffuse light as described in Section 
    \ref{Sec:dynamical_models}. The velocity dispersion profile of the GCs is also replotted for comparison 
    purposes. As can be seen, the velocity dispersion of the diffuse light and GCs are identical over the 
    common range of around 100 to 200 arcsec.
    }
   \label{fig:ngc3923_sigma}
\end{figure*}

Figure \ref{fig:ngc3923_sigma} presents the radial profile of the velocity dispersion of the NGC~3923 
diffuse light (green squares and stars) and total GC system (black line). The velocity dispersion of the 
GCs was estimated using the lowess estimator \citep{Gebhardt94,Gebhardt95}, which estimates the 
velocity variance at each datapoint by fitting a least squares linear fit to the individual 
points in a kernel (essentially in radius) around the point in question. Each 
datapoint in the fit is weighted by the inverse square of the distance 
between the point being fitted and the others within the kernel. The velocity dispersion is simply 
estimated as the square root of the variance measured by the lowess method. As a consistency check, 
this method was compared with the velocity dispersion estimations produced by a maximum likelihood 
method; for any reasonable bin size (minimum number of GCs per bin $>$ 15), the maximum likelihood 
method recovered velocity dispersions did not differ significantly from those found using the lowess 
method. 

The observation that the velocity dispersion profile of the NGC~3923 GCs is approximately constant, or 
slightly rising at larger radii, is a general finding in studies of the GCs of early type galaxies, as was 
discussed at length in Section \ref{Sec:Introduction}. 

Of particular interest in Figure \ref{fig:ngc3923_sigma} is the fact that the velocity dispersions of the 
integrated light of NGC~3923 determined from MOS slitlets using the method outlined in \cite{Norris08} 
(green stars) are consistent with the velocity dispersions of the GC system at the same radii. The 
implications of these results are best understood through the examination of simple dynamical 
models of NGC~3923.


\subsection{Dynamical Models}
\label{Sec:dynamical_models}

We use the axisymmetric, orbit-based modeling algorithm described in \cite{Gebhardt2000,Gebhardt03}, 
\cite{Thomas04DM,Thomas05DM} and \cite{Siopis09}. These models are based on the method of 
\cite{Schwarzschild79}, and similar models are presented in \cite{Richstone84}, \cite{Rix97}, \cite{Cretton99}, 
\cite{Valluri04}, \cite{Cappellari06}.

The modelling procedure consists of the following: (i) the surface-brightness distribution is deprojected 
into a 3d luminosity profile; (ii) a trial gravitational potential is calculated including the contribution from 
the stars and a dark matter halo; (iii) a large set of orbits is run in the trial potential; (iv) an orbit superposition 
model of the galaxy in the trial potential is constructed that matches the kinematical data subject to the 
constraints given by the 3d light profile and the trial potential; (v) steps (ii)-(iv) are repeated for a large 
variety of trial mass profiles. A $\chi^2$-analysis then determines the best-fit model and its uncertainties.

Due to the difference in the slopes of the profiles (surface brightness for stars and 
number for GCs), we model the stars and GCs separately, and then sum their $\chi^2$ as a measure 
of the quality of the fit for the given mass profile. Thus, the only addition to the above procedure is that, 
in step (iv), we fit a different set of orbital weights for stars and clusters. Each model is also constrained 
by the surface brightness/number density profile of the particular tracer. In this case we use 
the observed V-band surface brightness profile for the diffuse light of NGC~3923, and the observed total 
GC sample number density profile as shown in Figure \ref{fig:surface_brightness} and discussed in 
Section \ref{fig:surface_brightness}. Both stars and GCs are assumed to share the same 
flattening of $\epsilon$ = 0.4, consistent with the E4-5 classification and the measured
ellipticity of NGC~3923 found by \cite{Sikkema07}.

The NGC~3923 models use a polar geometric layout as follows. We use 20 radial bins and 5 angular 
bins for the spatial sampling, and 15 velocity bins. The radial spatial binning ranges from 1 arcsec in 
width at small radii up to 50 arcsec in the outer bins. The 5 angular spatial bins are equally spaced in 
sin($i$), where $i$ is measured from the major axis to the pole (i.e., angular bins are smaller closer to 
the major axis). The velocity bins are 100~kms$^{-1}$ wide. The average number of orbits per model is 
40,000. The orbital sampling follows the design in \cite{Thomas04DM,Thomas05DM}.

We fit for three parameters that determine the mass profile; these are the stellar M/L assumed constant 
with radius, the dark halo scale radius, and the dark halo density at $R=0$. 
To represent the dark matter profile, we use a spherical cored logarithmic profile given as:
\\
\begin{eqnarray}
\rho_{\rm DM}(r) = \frac{v_{\rm c}^2}{4\pi G} \frac{3r_{\rm c}^2 + r^2}{(r_{\rm c}^2 + r^2)^2}.
\end{eqnarray}
\\
\noindent where $v_{\rm c}$ is circular velocity at the core radius, $r_{\rm c}$ \citep{Binney_Tremaine}.

\cite{Thomas05DM}, \cite{Forestell10}, and \cite{Murphy11} find that a cored logarithmic profile 
provides a better fit or at least as good a fit as an NFW profile. Given 
that we do not have kinematic data at large radii in large enough quantity to explore the various dark halo 
parameterizations, we only consider the one model.  In all models, we include a black hole of mass 
1$\times$10$^8$~M$_{\odot}$, which is the mass derived from its correlation with velocity dispersion
\citep{Gultekin10}; the value of the black hole mass will not affect the dark matter profile and 
we only include it for completeness.

\begin{figure} 
   \centering
   \begin{turn}{0}
   \includegraphics[scale=0.85]{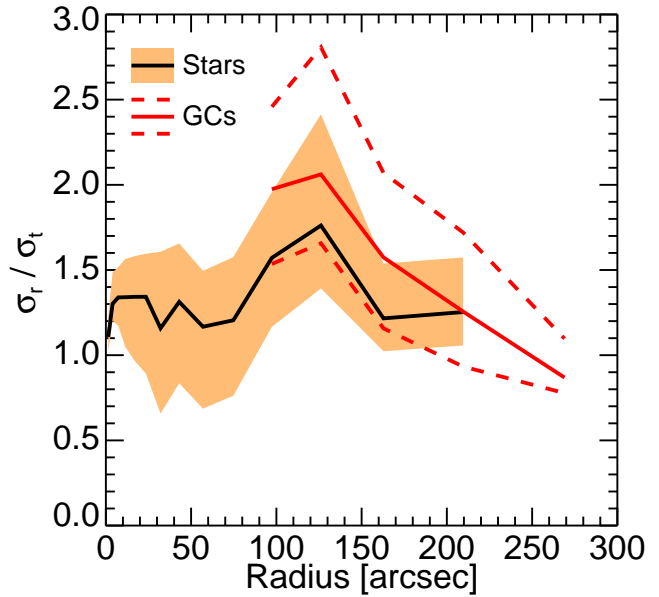}
   \end{turn}
    \caption[Orbits]
    {The ratio of the radial to tangential velocity dispersion as determined using
    our dynamical modelling procedure outlined in Section \ref{Sec:dynamical_models}.
    The solid black line shows the behaviour of the diffuse stellar light of NGC~3923,
    with the orange shaded region displaying the $\pm$1$\sigma$ errors. The solid red
    line shows the behaviour of the GCs of NGC~3923, with the dashed red lines denoting
    the $\pm$1$\sigma$ errors.}
   \label{fig:ngc3923_orbits}
\end{figure}

\begin{figure*} 
   \centering
   \begin{turn}{0}
   \includegraphics[scale=0.9]{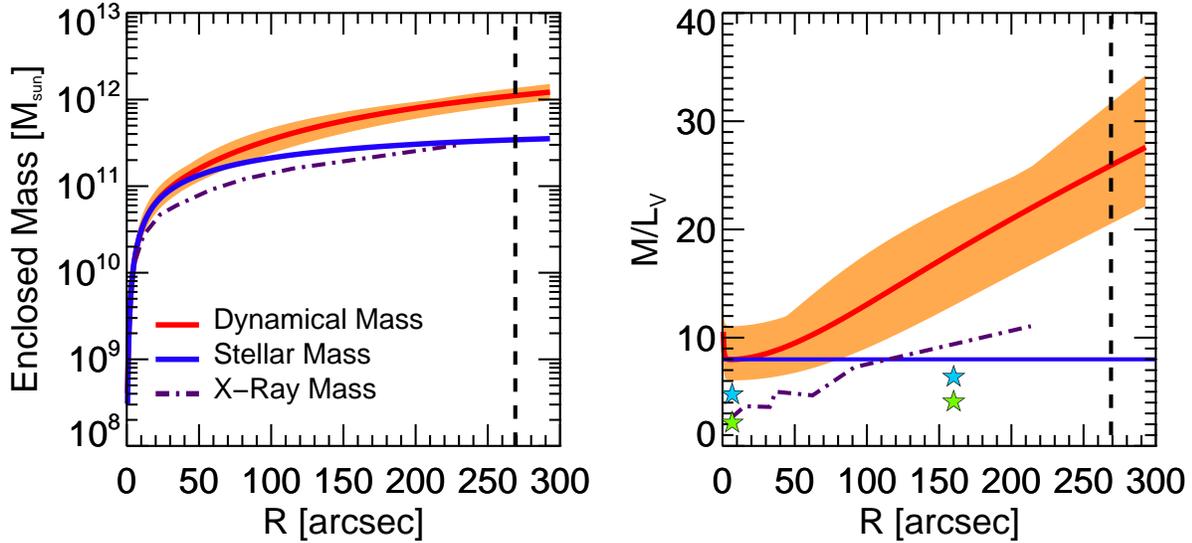}
   \end{turn}
    \caption[Mass Profile]
    {\textbf{Left~Panel:} Enclosed mass profiles, for the dynamical mass (red line), stellar mass (blue line), 
    and X-ray mass profile from \cite{Fukazawa06}. The vertical dashed line shows the limit of the binned 
    GC velocity data.
    \textbf{Right~Panel:} Mass-to-light ratio profile. The apparent turn up of the model within 10 arcsec is due 
    to the inclusion in the model of a nominal black hole of mass 1$\times$10$^8$~M$_{\odot}$.  The star symbols 
    represent SSP model predictions for M/L$_{\rm V}$ for the stellar population given the age and metallicity 
    measured for the population either in \citet{Norris08} (larger radii point) or by \citet{Thomas05} (small radii 
    point). The SSP models come from \citet{Maraston05} for Salpeter IMF (upper/blue star symbols) and Kroupa 
    IMF (lower/green star symbols).}
   \label{fig:ngc3923_mass_profile}
\end{figure*}

The results of this fitting procedure are displayed in Figures \ref{fig:ngc3923_sigma}, \ref{fig:ngc3923_orbits} and 
\ref{fig:ngc3923_mass_profile}. The best fit M/L$_{\rm V}$ in the central region is 8, identical to that 
found for NGC~4649 by \citet{Bridges06} and \citet{Shen10}, and close to the M/L$_{\rm V}$ of 6 
found for NGC~3379 by \citet{Pierce3379}. Mass-to-light ratios as high as these, while large, are 
only moderately inconsistent with the predictions of stellar population synthesis models for such old 
stellar populations (as the star symbols in the right panel of Figure \ref{fig:ngc3923_mass_profile} demonstrate).

In the upper panel of Figure~\ref{fig:ngc3923_sigma} the white and blue dashed line displays the 
best-fit model GC velocity dispersion profile. The fact that a model constrained by both the diffuse light 
velocity dispersion, and the GC kinematics (plus their respective number density profiles), is still able to 
successfully reproduce the observed GC velocity dispersion profile is reassuring. 

In addition to determining the best-fit mass profile, orbit-based models provide a measure 
of the internal orbital structure. Figure \ref{fig:ngc3923_orbits} presents the ratio of the radial to tangential 
internal velocity dispersion. We present this ratio for both the diffuse light (solid black line), and GCs (solid red line) 
of NGC~3923. In general, the large uncertainties prevent any strong conclusions on the orbital behaviour 
of the GCs and diffuse light. However, to first order the profiles of both appear to be mildly radially biased 
at all but the largest radii. Additionally, the GCs and diffuse light of NGC~3923 appear to show similar 
behaviour over the region of overlap. Beyond a radius of around 130$^{\prime\prime}$ both the GCs and 
diffuse light appear to become more isotropic ($\sigma_{\rm r}$/$\sigma_{\rm t}$=1).

In the left panel of Figure \ref{fig:ngc3923_mass_profile} the inferred enclosed mass profiles are displayed.
Of immediate interest is the fact that the mass profile determined from X-ray observations by 
\cite{Fukazawa06} is around a factor of two less than our total dynamical mass profile, though with similar 
slope at larger radii. In fact, the X-ray mass profile is lower than even our stellar only model (blue line).
Therefore the disagreement cannot be due to differences in the derived DM halos of the models. However,
the stellar mass and the X-ray mass profile could be made to agree, if a less extreme M/L$_{\rm V}$ of 
around 5 was chosen (which would also be more consistent with stellar population models), 
though this would then be inconsistent with the observed velocity dispersion of the integrated light and GCs 
of NGC~3923.

Discrepancies of similar magnitude between X-rays and GCs have been found in 
NGC~4636 \citep{Johnson09}, NGC~1407 \citep{Romanowsky09}, and NGC~4649 \citep{Shen10}. 
In fact disagreements between different dynamical tracers within the same galaxy are a general
result of studies of galaxy dynamics. In particular, PNe commonly indicate the presence of 
little or no DM in galaxies in which other tracers detect significant quantities e.g. in NGC~821 
PNe \citep{Romanowsky03,Teodorescu10} indicate little or no DM present, while stellar 
kinematics \citep{Weijmans09,Forestell10} indicate between 13 and 18$\%$ of the mass of 
NGC~821 is DM within R$_{\rm e}$. As all dynamical tracers have their
own strengths and weaknesses, only by combining several different tracers within a galaxy can we
hope to achieve robust results.

The best-fit parameters to our cored logarithmic profile
are $v_{\rm c}$=450~$\pm$~50~kms$^{-1}$, and r$_{\rm c}$=11~$\pm$~4~kpc.
The best-fit model has $\chi^2$=22. For the stellar kinematics, we only use the 
dispersions measured in 13 circular apertures at various radii from 1.5\arcsec to 200\arcsec. As discussed 
previously, we do not find significant rotation. For the globular clusters we use three spatial bins, and for 
each bin we consider both the velocity and velocity dispersion. Thus, the total degrees of freedom is 19. 
The reduced $\chi^2$ of 1.16 suggest that the dynamical model well represents the data. For the uncertainties, 
we rely on those models within a set $\Delta\chi^2$ value. We report 68\% confidence values, where we 
have marginalized over the other parameters; thus, the 1-sigma values in this case come from $\Delta\chi^2=1$. 
The confidence bands in Fig.~\ref{fig:ngc3923_mass_profile} represent the range in models that are within 
this $\chi^2$.

Our best fit cored logarithmic model returns an enclosed dark matter fraction of 17.5$^{+7.3}_{-4.5}$$\%$ 
within R$_{\rm e}$, which rises to 41.2$^{+18.2}_{-10.6}$$\%$ within 2R$_{\rm e}$. At the limit of our 
kinematic coverage (370 arcsec, 6.9 R$_{\rm e}$) dark matter comprises 75.6$^{+15.4}_{-16.8}$$\%$ of 
the mass of NGC~3923. This distribution of dark matter is similar to that measured in other massive early types 
using stellar absorption line kinematics or dynamical tracers, e.g. M87 \citep[17.2$\%$ DM within R$_{\rm e}$,][]{Murphy11},  
NGC~821 \citep[13 - 18$\%$ DM within R$_{\rm e}$,][]{Weijmans09,Forestell10}, 
NGC~4594 \citep[19$\%$ DM within R$_{\rm e}$,][]{Bridges07},
or 17 early-type galaxies in the Coma 
cluster \citep[10-50$\%$ DM within R$_{\rm e}$,][]{Thomas07DM}. It is however slightly inconsistent with 
other studies which have found larger DM fractions using either SAURON observations
\citep[median DM fraction 30$\%$ within R$_{\rm e}$,][]{Cappellari06} or observations of strong
gravitation lenses \citep[38$\%$ DM within R$_{\rm e}$,][]{Bolton08}.

The right panel of Figure \ref{fig:ngc3923_mass_profile} shows the behaviour of the M/L$_{\rm V}$ profile 
of NGC~3923. At the limit of the binned data (269 arcsec) the inferred M/L$_{\rm V}$ of 25.9 is inconsistent 
with the constant M/L$_{\rm V}$ (of 8) at the 3.3$\sigma$ level. This again confirms the existence of 
significant amounts of DM in NGC~3923.


\section{Discussion}
An examination of the GC radial velocities has not detected any significant rotation in the NGC~3923 GC system. This 
finding means that the GCs display similar behaviour to the integrated light of NGC~3923 which also has 
little or no rotation \citep[e.g][]{Carter98,Koprolin2000,Norris08}. This lack of rotation in the GC systems of 
massive early-type galaxies is not uncommon, with little or no rotation observed in the GC systems of 
NGC~4649 \citep{Bridges06}, NGC~1399 \citep{Richtler04,Schuberth10}, NGC~3379 \citep{Bergond06},
NGC~4472 \citep{Zepf00,Cote03}, and NGC~4594 \citep{Bridges07}. Some GC systems however do display 
significant rotation, for example NGC~5128 \citep{Woodley07} and M87 \citep{Cote01}. The cause of this variation 
may be partly due to projection effects. Nevertheless the current observations do not fit comfortably with 
theoretical simulations of galaxy formation. These simulations tend to predict that significant amounts of angular 
momentum should be found in the outer regions of galaxies, especially in cases where the galaxies formed in 
major merger events in the relatively more recent past \citep[i.e. z $<$ 3,][]{vitvitska02}. 

To explain this disagreement it may be possible to call on angular momentum transport to move the angular 
momentum beyond the radii to which current GC studies are presently sensitive; only an extension of these 
studies to even larger galactocentric radii would be able to determine if this is indeed the case. This would 
most likely prove impossible for all but the most massive galaxies with the richest GC systems, as smaller 
galaxies simply would not have enough GCs at sufficiently large radii to make such a measurement possible. 

Turning to the velocity dispersions of the GCs and integrated light of NGC~3923 reveals that these also display 
indistinguishable behaviours over the range where they can both be measured (see Fig. \ref{fig:ngc3923_sigma}). 
It is also intriguing that the value of the observed constant velocity dispersion of $\sim$200~kms$^{-1}$ is 
indistinguishable from the dispersion measured for the NGC~3923 group as a whole by \cite{Brough06},
using the redshifts of all 30 group members.
It should be noted however, that there are indications that the integrated light velocity dispersion may in fact 
be closer to that of the red GC subsample, rather than the system as a whole. The weighted mean of the 
integrated light velocity dispersion for the four largest radii points is 198 $\pm$ 9~kms$^{-1}$, consistent with 
the 200 $\pm$ 22~kms$^{-1}$ found for the red GCs (Figure \ref{fig:ngc3923_hist}), but marginally inconsistent 
with the velocity dispersion of the blue GCs of 272 $\pm$ 34~kms$^{-1}$. This is compatible with the halo-GCs
connection discussed in \citet{Forte07,Forte09}, in the sense that most of the galaxy halo light is contributed
by the diffuse stellar population associated with these red subpopulation clusters. Nonetheless 
this is not to say that the kinematics of red GCs alone would be a preferable tracer of the mass profile of 
NGC~3923. In principle, with a sufficient sample size, and hence the correct number profile and orbital 
information red and blue GCs should give the same mass profile.

Because of the difficulty of measuring stellar absorption line kinematics to large radii in early type galaxies,
very few studies have been able to directly compare GC and integrated light kinematics over the same radii. 
However, \cite{Schuberth10} found that the red GCs and integrated light of NGC~1399 likewise display
similar velocity dispersions over their common range of study. If this finding proves to be a general result
of studies of GC kinematics it will provide a very important clue to the formation history of galaxies.
Such a close correspondence in kinematics, combined with the observed similarities in distribution
(see Figure \ref{fig:surface_brightness} and Section \ref{Sec:spatial_dist}), colours \citep[see][]{ForbesForte01,Sikkema06},
and stellar populations \citep[see][]{Norris08}, indicate a common formation for the red GCs and 
diffuse light of NGC~3923.

The observation that the velocity dispersions of the GCs and integrated light of NGC~3923 are constant
at larger radii, despite the rapidly declining number density/surface brightness, is already strong 
evidence for the existence of significant amounts of DM in this galaxy. Our orbit-based models
confirm that NGC~3923 has dark matter fractions consistent with those found in other studies of
early type galaxies using integrated light or GC dynamical tracers. 

The observation that our estimates of dynamical mass (and hence M/L$_{\rm V}$) are larger than those made 
with X-ray observations is intriguing. \citet{Shen10} find discrepancies of similar magnitude and in the
same sense in their study of NGC~4649, as do \citet{Murphy11} for their study of the stellar kinematics
of M87. In contrast, \citet{Johnson09} find disagreement in both senses
\textit{in the same galaxy}, NGC~4636, where within 10~kpc they find the GCs indicate $\sim$twice as much 
mass as the X-rays, between 10 and 30~kpc GCs and X-rays agree, and then beyond 30~kpc X-rays indicate 4-5 times
as much mass as the GCs. Like \citet{Shen10} we suggest that the observed 
discrepancy may indicate problems with the X-ray modelling. Perhaps other forms of non-thermal pressure 
support, such as magnetic fields, microturbulence, or cosmic-rays are present in the X-ray gas, but are 
not yet adequately treated in the X-ray mass determination. Additionally, the assumption that the X-ray 
emitting gas is in hydrostatic equilibrium may need to be relaxed. This is a particular concern in disturbed 
galaxies such as NGC~3923, where its obvious shell structures may imply a relatively recent merger event. 
These results act to reinforce the fact that only by combining as many mass tracers as possible within 
individual galaxies is it possible to overcome each tracers shortcomings and arrive at a robust estimate of 
galaxy dynamics and DM content.

\section{Conclusions}

This paper has presented Gemini/GMOS spectra of 79 GCs associated with NGC~3923 measured using 
MOS and N$\&$S techniques. Their radial velocities have been combined with GC number densities 
profiles, stellar integrated light kinematics, and stellar surface brightness profiles and used to constrain 
the DM content of the NGC~3923. Our main conclusions are:

\begin{enumerate}

\item	There is no significant evidence for rotation in any of the GC sub-samples (total, blue, or red)
	out to the largest projected radius measured (6.9R$_{\rm e}$). \\
	
\item	In common with other studies of early type galaxies \citep[e.g.][]{Schuberth10,Park10} the blue 
	GC subsample appears to have a higher dispersion (272 $\pm$ 34 kms$^{-1}$), than the red GC 
	subsample (200 $\pm$ 22 kms$^{-1}$). But with our current sample size this difference is not yet 
	statistically significant. \\

\item	The GC velocity dispersion profile for the total GC sample is consistent with being constant with 
	radius. \\
	
\item	The velocity dispersion of the integrated light of NGC~3923 at large radii is also constant, and is 
	consistent with the GC velocity dispersion profile at the same radii. \\
	
\item We find some evidence that the diffuse light and GCs of NGC~3923 have
	radially biased orbits within $\sim$130$^{\prime\prime}$. \\

\item	By applying axisymmetric orbit-based models to the GC and integrated light data we find that 
	NGC~3923 is 17.5$^{+7.3}_{-4.5}$$\%$ DM within R$_{\rm e}$, 41.2$^{+18.2}_{-10.6}$$\%$ 
	within 2R$_{\rm e}$, and 75.6$^{+15.4}_{-16.8}$$\%$ within 6.9R$_{\rm e}$. These
	DM mass fractions for a group elliptical are indistinguishable from those found using similar 
	methods for ellipticals in environments from isolated (NGC821) to cD (M87), perhaps indicating
	the universal nature of the DM profile.\\
	
\item The total dynamical mass within the radius of our last dynamical tracer (6.9R$_{\rm e}$) is found 
	to be 1.5$^{+0.4}_{-0.25}$$\times$10$^{12}$~M$_\odot$. \\	
	
\item	The inferred GC+integrated light dynamical mass and X-ray mass profiles are discrepant. The X-ray 
	mass profile is around 50$\%$ of that of the dynamical mass.

\end{enumerate}

\section{Acknowledgements}

The authors would like to thank David Fisher for providing
the deep V-band surface brightness profile of NGC~3923
used in this paper and Yasushi Fukazawa for supplying
the B band M/L profile of NGC~3923 as measured from
X-ray observations.

MAN acknowledges financial support from the STFC. 
Support for Program number HST-AR-12147.01-A was provided by 
NASA through a grant from the Space Telescope Science 
Institute, which is operated by the Association of Universities 
for Research in Astronomy, Incorporated, under NASA 
contract NAS5-26555.

SEZ  acknowledges support from NSF award AST-0406891.
DF acknowledges support from the ARC. FF acknowledges financial
support from the Agencia de Promoci\'{o}n Cientifica y Tecnol\'{o}gica (BID AR PICT 885).

We are thankful for the facilities at the Texas Advanced Computing 
Center at The University of Texas at Austin, which has allowed access 
to over 5000 node computers where we ran all of the models. KG 
acknowledges support from NSF-0908639.

This paper makes use of data obtained as part of Gemini Observtory 
programs  GS-2004A-Q-9 and GS-2011A-Q-13. 	

Based on observations obtained at the Gemini Observatory, which is 
operated by the Association of Universities for Research in Astronomy, Inc., 
under a cooperative agreement with the NSF on behalf of the Gemini 
partnership: the National Science Foundation (United States), the Science 
and Technology Facilities Council (United Kingdom), the National Research 
Council (Canada), CONICYT (Chile), the Australian Research Council (Australia), 
MinistŽrio da Cincia, Tecnologia e Inova‹o (Brazil) and Ministerio de Ciencia, 
Tecnolog'a e Innovaci—n Productiva (Argentina).

 \begin{appendix}
  
\bibliographystyle{mn2e}
\bibliography{references}
\label{lastpage}

\appendix
 \begin{table*}
\begin{center}
\caption[NGC~3923 GCs - 1]{Confirmed NGC~3923 GCs. ID, RA, Dec, X, Y, g, r, i, velocity and velocity errors
for all 79 confirmed NGC~3923 GCs/UCDs. The ID refers to the GCs position in the photometry table of \citet{Faifer11}.
The X and Y positions refer to positions relative to the galaxy centre. Name denotes the ID given to objects studied 
in \cite{Norris08} or \cite{Norris&Kannappan11}.}
\begin{tabular}{lcccccccccc} \hline
 ID	&	RA 		& Dec. 	& 	X		&	 Y		& 	g$^\prime_0$		&	 r$^\prime_0$ 	& i$^\prime_0$ 		& V 		& Name	 \\
	&	(J2000)	& (J2000)	&	(arcmin)	& (arcmin)	&	(mag)        &   (mag)   & (mag)    & (kms$^{-1}$)  	&	\\
\hline
       26 & 177.70500 & -28.78370 &  2.74 &  1.35 & 22.52 & 22.05 & 21.85 & 1586.5 $\pm$ 66.0 & NSB228 \\ 
       35 & 177.70690 & -28.82370 &  2.64 & -1.05 & 21.43 & 20.81 & 20.53 & 1590.8 $\pm$ 35.9 &  \\ 
       39 & 177.70700 & -28.81640 &  2.63 & -0.61 & 22.35 & 21.68 & 21.38 & 1726.6 $\pm$ 28.9 & NSB197 \\ 
       55 & 177.78160 & -28.77420 & -1.29 &  1.92 & 21.31 & 20.78 & 20.59 & 2295.2 $\pm$ 47.3 & NSB65 \\ 
       58 & 177.70860 & -28.83960 &  2.55 & -2.00 & 22.68 & 22.15 & 21.98 & 2203.2 $\pm$ 74.2 & NSB253 \\ 
       90 & 177.71090 & -28.78090 &  2.43 &  1.52 & 23.08 & 22.61 & 22.35 & 1851.2 $\pm$ 69.2 & NSB394 \\ 
     103 & 177.77170 & -28.81650 & -0.77 & -0.62 & 20.61 & 20.04 & 19.80 & 2307.9 $\pm$ 34.8 & NGC3923-UCD3 \\ 
     128 & 177.78560 & -28.77520 & -1.50 &  1.86 & 21.76 & 21.27 & 21.07 & 2144.1 $\pm$ 45.0 & NSB104 \\ 
     158 & 177.78340 & -28.81080 & -1.38 & -0.27 & 22.69 & 21.99 & 21.63 & 1809.0 $\pm$ 31.7 &  \\ 
     162 & 177.78300 & -28.76930 & -1.36 &  2.22 & 22.12 & 21.52 & 21.23 & 1810.9 $\pm$ 26.9 & NSB167  \\ 
     176 & 177.78150 & -28.77910 & -1.28 &  1.63 & 21.22 & 20.60 & 20.32 & 1995.9 $\pm$ 50.4 &  \\ 
     193 & 177.77960 & -28.83760 & -1.18 & -1.88 & 21.80 & 21.13 & 20.78 & 2046.0 $\pm$ 32.5 & NSB108  \\ 
     206 & 177.77880 & -28.82970 & -1.14 & -1.41 & 22.73 & 22.20 & 21.99 & 1752.7 $\pm$ 45.4 &  \\ 
     214 & 177.74260 & -28.84840 &  0.76 & -2.53 & 22.86 & 22.13 & 21.81 & 1925.2 $\pm$ 55.0 & NSB333 \\ 
     243 & 177.73280 & -28.80520 &  1.28 &  0.06 & 19.99 & 19.42 & 19.17 & 1478.0 $\pm$ 28.6 & NGC3923-UCD2  \\ 
     251 & 177.77530 & -28.80570 & -0.96 &  0.03 & 21.36 & 20.60 & 20.22 & 1790.4 $\pm$ 28.4 &  \\ 
     275 & 177.77380 & -28.83190 & -0.88 & -1.54 & 22.88 & 22.13 & 21.69 & 1638.9 $\pm$ 40.9 & NSB332  \\ 
     277 & 177.77380 & -28.82810 & -0.88 & -1.31 & 22.68 & 22.02 & 21.67 & 1945.0 $\pm$ 64.2 &  \\ 
     312 & 177.77010 & -28.76440 & -0.69 &  2.51 & 21.84 & 21.22 & 20.88 & 1555.1 $\pm$ 35.7 & NSB110  \\ 
     318 & 177.76990 & -28.78640 & -0.67 &  1.19 & 23.03 & 22.39 & 22.06 & 1555.8 $\pm$ 39.0 &  \\ 
     336 & 177.76700 & -28.80560 & -0.52 &  0.04 & 19.53 & 18.88 & 18.58 & 2114.7 $\pm$ 29.9 & NGC3923-UCD1 \\ 
     344 & 177.76680 & -28.76360 & -0.51 &  2.56 & 22.49 & 22.01 & 21.82 & 1805.4 $\pm$ 53.1 & NSB221 \\ 
     365 & 177.76470 & -28.83240 & -0.40 & -1.57 & 22.52 & 21.81 & 21.47 & 1936.0 $\pm$ 56.0 & NSB225  \\ 
     375 & 177.76410 & -28.79420 & -0.37 &  0.72 & 21.23 & 20.52 & 20.15 & 1438.4 $\pm$ 27.7 &  \\ 
     380 & 177.77950 & -28.82690 & -1.18 & -1.24 & 21.44 & 20.97 & 20.78 & 1159.5 $\pm$ 50.2 &  \\ 
     396 & 177.76120 & -28.78020 & -0.22 &  1.56 & 22.70 & 22.10 & 21.81 & 1190.8 $\pm$ 44.5 &  \\ 
     419 & 177.75850 & -28.82620 & -0.07 & -1.20 & 21.88 & 21.40 & 21.20 & 1564.0 $\pm$ 60.2 &  \\ 
     453 & 177.75500 & -28.76550 &  0.11 &  2.45 & 23.46 & 22.86 & 22.42 & 1618.9 $\pm$ 61.2 & NSB513  \\ 
     458 & 177.75470 & -28.81330 &  0.13 & -0.42 & 21.60 & 21.12 & 20.93 & 1617.4 $\pm$ 62.7 &  \\ 
     484 & 177.75260 & -28.76210 &  0.24 &  2.65 & 23.26 & 22.65 & 22.37 & 2114.4 $\pm$ 38.9 & NSB450  \\ 
     492 & 177.75160 & -28.78010 &  0.29 &  1.57 & 22.22 & 21.55 & 21.22 & 1760.6 $\pm$ 15.2 &  \\ 
     502 & 177.75060 & -28.80160 &  0.34 &  0.28 & 21.56 & 20.76 & 20.36 & 1627.5 $\pm$ 26.6 &  \\ 
     503 & 177.75060 & -28.78290 &  0.34 &  1.40 & 22.56 & 21.96 & 21.57 & 1807.9 $\pm$ 29.6 & NSB232  \\ 
     530 & 177.72510 & -28.80540 &  1.68 &  0.05 & 21.28 & 20.68 & 20.42 & 1764.7 $\pm$ 32.0 & NSB64 \\ 
     534 & 177.74800 & -28.80790 &  0.48 & -0.10 & 21.56 & 20.88 & 20.54 & 2148.3 $\pm$ 26.1 &  \\ 
     535 & 177.74780 & -28.83600 &  0.49 & -1.78 & 22.97 & 22.27 & 21.88 & 1457.0 $\pm$ 70.4 & NSB360  \\ 
     552 & 177.74640 & -28.76710 &  0.56 &  2.35 & 23.33 & 22.60 & 22.26 & 1749.1 $\pm$ 74.1 & NSB492 \\ 
     557 & 177.74590 & -28.83100 &  0.59 & -1.48 & 23.03 & 22.35 & 22.07 & 1621.8 $\pm$ 73.0 &  \\ 
     593 & 177.74050 & -28.79270 &  0.87 &  0.81 & 21.87 & 21.27 & 20.99 & 2114.6 $\pm$ 52.7 &  \\ 
     595 & 177.74140 & -28.77960 &  0.82 &  1.60 & 22.83 & 22.10 & 21.74 & 1796.2 $\pm$ 55.6 &  \\ 
     601 & 177.74040 & -28.77930 &  0.88 &  1.62 & 22.79 & 22.20 & 21.93 & 1666.4 $\pm$ 42.5 & NSB311  \\ 
     617 & 177.73830 & -28.80010 &  0.99 &  0.37 & 21.87 & 21.18 & 20.86 & 1949.9 $\pm$ 24.2 &  \\ 
     621 & 177.73840 & -28.83810 &  0.98 & -1.91 & 22.61 & 22.06 & 21.86 & 1840.7 $\pm$ 48.2 & NSB247  \\ 
     669 & 177.73420 & -28.76740 &  1.20 &  2.33 & 21.62 & 21.03 & 20.72 & 1784.4 $\pm$ 51.1 & NSB86 \\ 
     683 & 177.73260 & -28.77300 &  1.29 &  2.00 & 22.80 & 22.08 & 21.67 & 1755.7 $\pm$ 53.9 &  \\ 
     692 & 177.73240 & -28.78010 &  1.30 &  1.57 & 23.36 & 22.75 & 22.42 & 1622.1 $\pm$ 48.6 & NSB498  \\ 
     713 & 177.73050 & -28.77580 &  1.40 &  1.83 & 22.83 & 22.30 & 22.08 & 2038.4 $\pm$ 88.5 & NSB322 \\ 
     725 & 177.72900 & -28.80110 &  1.48 &  0.31 & 21.72 & 21.08 & 20.77 & 1664.7 $\pm$ 24.9 & NSB99 \\ 
     760 & 177.69530 & -28.82620 &  3.25 & -1.20 & 20.50 & 19.89 & 19.62 & 1714.4 $\pm$ 26.3 &  \\ 
     779 & 177.72420 & -28.79050 &  1.73 &  0.95 & 21.42 & 20.85 & 20.61 & 1861.3 $\pm$ 43.8 &  \\ 
     791 & 177.72330 & -28.76960 &  1.78 &  2.20 & 21.73 & 21.14 & 20.85 & 1733.2 $\pm$ 26.3 & NSB106  \\ 
     824 & 177.79280 & -28.80270 & -1.88 &  0.21 & 21.62 & 21.03 & 20.79 & 1756.4 $\pm$ 42.5 &  \\ 
     847 & 177.68510 & -28.85610 &  3.78 & -2.99 & 21.89 & 21.45 & 21.06 & 2024.5 $\pm$ 68.1 &  \\ 
     859 & 177.71780 & -28.83340 &  2.06 & -1.63 & 21.66 & 21.00 & 20.68 & 1970.9 $\pm$ 59.7 &  \\ 
     872 & 177.71640 & -28.80990 &  2.14 & -0.22 & 21.87 & 21.27 & 21.02 & 1762.7 $\pm$ 26.2 & NSB112  \\ 
     903 & 177.71360 & -28.84480 &  2.29 & -2.31 & 22.70 & 21.98 & 21.61 & 1796.6 $\pm$ 57.9 & NSB279 \\ 
     940 & 177.80100 & -28.80420 & -2.31 &  0.12 & 21.26 & 20.74 & 20.55 & 1494.4 $\pm$ 56.9 &  \\ 
     952 & 177.78790 & -28.80630 & -1.62 & -0.00 & 21.66 & 21.00 & 20.67 & 1738.2 $\pm$ 25.3 & NSB93 \\ 
 \hline
\label{tab:ngc3923gckin}
\end{tabular}
\end{center}
\end{table*}

\begin{table*}
\begin{center}
\caption[NGC~3923 GCs]{Table \ref{tab:ngc3923gckin} continued.}
\begin{tabular}{lccccccccc} \hline
 ID	&	RA 		& Dec. 	& 	X		&	 Y		& 	g$^\prime_0$ 		&	 r$^\prime_0$ 	& i$^\prime_0$ 		& V 		& Name	 \\
	&	(J2000)	& (J2000)	&	(arcmin)	& (arcmin)	&	(mag)        &   (mag)   & (mag)    & (kms$^{-1}$)  	&	\\
\hline
      977 & 177.79550 & -28.76420 & -2.02 &  2.52 & 22.35 & 21.82 & 21.64 & 2219.1 $\pm$ 65.9 &  \\ 
    1030 & 177.66740 & -28.82500 &  4.71 & -1.12 & 20.75 & 20.21 & 19.95 & 1517.0 $\pm$ 32.1 &  \\ 
    1036 & 177.79440 & -28.79320 & -1.96 &  0.78 & 22.04 & 21.41 & 21.12 & 1811.5 $\pm$ 44.2 &  \\ 
    1107 & 177.65980 & -28.86380 &  5.11 & -3.45 & 21.49 & 20.97 & 20.76 & 1354.7 $\pm$ 62.0 &  \\ 
    1252 & 177.82400 & -28.78750 & -3.52 &  1.12 & 21.97 & 21.37 & 21.16 & 1731.1 $\pm$ 27.1 &  \\ 
    1297 & 177.70040 & -28.78040 &  2.98 &  1.55 & 22.23 & 21.58 & 21.26 & 1922.9 $\pm$ 24.3 &  \\ 
    1310 & 177.69830 & -28.84090 &  3.09 & -2.08 & 21.28 & 20.67 & 20.40 & 2109.3 $\pm$ 23.5 &  \\ 
    1320 & 177.69750 & -28.78430 &  3.13 &  1.32 & 22.25 & 21.50 & 21.11 & 1988.6 $\pm$ 31.4 &  \\ 
    1356 & 177.69350 & -28.81410 &  3.34 & -0.47 & 22.02 & 21.36 & 21.06 & 1903.5 $\pm$ 39.4 &  \\ 
    1437 & 177.68240 & -28.84210 &  3.93 & -2.15 & 21.53 & 20.99 & 20.68 & 1847.5 $\pm$ 23.1 &  \\ 
    1524 & 177.67090 & -28.83400 &  4.53 & -1.66 & 22.32 & 21.82 & 21.61 & 1300.5 $\pm$ 28.3 &  \\ 
    1535 & 177.66890 & -28.79070 &  4.64 &  0.93 & 22.82 & 22.09 & 21.72 & 1739.8 $\pm$ 31.6 &  \\ 
    1613 & 177.65850 & -28.80990 &  5.18 & -0.22 & 21.54 & 20.98 & 20.74 & 1782.5 $\pm$ 63.9 &  \\ 
    1653 & 177.73540 & -28.84380 &  1.14 & -2.25 & 22.28 & 21.40 & 21.19 & 1769.9 $\pm$ 20.8 &  \\ 
    1694 & 177.82870 & -28.75210 & -3.77 &  3.25 & 21.78 & 21.20 & 20.92 & 1734.0 $\pm$ 40.1 &  \\ 
    1797 & 177.75200 & -28.72930 &  0.27 &  4.62 & 22.26 & 21.59 & 21.37 & 2167.9 $\pm$ 39.9 &  \\ 
    1867 & 177.76790 & -28.74720 & -0.57 &  3.54 & 22.02 & 21.45 & 21.25 & 1635.2 $\pm$ 51.8 &  \\ 
    1903 & 177.77520 & -28.72790 & -0.95 &  4.70 & 23.10 & 22.40 & 22.12 & 1992.7 $\pm$ 55.5 &  \\ 
    1919 & 177.77810 & -28.73110 & -1.11 &  4.51 & 21.83 & 21.18 & 20.90 & 2039.1 $\pm$ 33.3 &  \\ 
    2153 & 177.81160 & -28.79920 & -2.87 &  0.42 & 23.24 & 22.54 & 22.21 & 1975.5 $\pm$ 39.0 &  \\ 
    2223 & 177.81990 & -28.77930 & -3.30 &  1.61 & 22.33 & 21.63 & 21.28 & 2064.4 $\pm$ 15.7 &  \\ 
\hline
\label{tab:ngc3923gckin2}
\end{tabular}
\end{center}
\end{table*}

\end{appendix}

\end{document}